\documentclass[iop]{emulateapj}
\usepackage{graphicx}
\usepackage{epsfig}

\shorttitle{FRB/GRB cosmography} \shortauthors{Gao, Li \& Zhang} \slugcomment{}

\begin{document}

\title{Fast Radio Burst/Gamma-Ray Burst Cosmography}

\author{He Gao$^{1}$, Zhuo Li$^{2,3}$, Bing Zhang$^{1,2,3}$}
\affil{$^1$Department of Physics and Astronomy, University of Nevada Las Vegas, NV 89154, USA;gaohe@physics.unlv.edu;~zhang@physics.unlv.edu\\
  $^2$Department of Astronomy, School of Physics, Peking University, Beijing 100871, China; zhuo.li@pku.edu.cn\\
  $^3$Kavli Institute of Astronomy and Astrophysics, Peking University, Beijing 100871, China\\}

\begin{abstract}
Recently, both theoretical arguments and observational evidence suggested that a small fraction of fast radio bursts (FRBs) could be associated with gamma-ray bursts (GRBs). If such FRB/GRB association systems are commonly detected in the future, the combination of dispersion measure (DM) derived from FRBs and redshifts derived from GRBs makes these systems a plausible tool to conduct cosmography. We quantify uncertainties in deriving the redshift-dependent ${\rm DM_{\rm IGM}}$ as a function of $z$, and test how well dark energy models can be constrained with Monte Carlo simulations. We show that with potentially several 10s of FRB/GRB systems detected in a decade or so, one may reach reasonable constraints on $w$CDM models. When combined with SN Ia data, unprecedented constraints on dark energy equation of state may be achieved, thanks to the prospects of detecting FRB/GRB systems at relatively high redshifts. The ratio between the mean value $\left< {\rm {DM}_{\rm IGM}} (z)\right>$ and luminosity distance ($D_{\rm L} (z)$) is insensitive to dark energy models. This gives the prospects of applying SN Ia data to calibrate $\left< {\rm {DM}_{\rm IGM}} (z)\right>$ using a relatively small sample of FRB/GRB systems, allowing a reliable constraint on the baryon inhomogeneity distribution as a function of redshift. The methodology developed in this paper can also be applied, if the FRB redshifts can be measured by other means. Some caveats of putting this method into practice are also discussed.
\end{abstract}

\section{Introduction}

The nature of late time cosmic acceleration is a deep mystery in cosmology and fundamental physics, which could be explained by introducing an exotic form of energy content with negative pressure, dubbed dark energy. Cosmological parameters have been measured via various standard candles or rulers, such as Type Ia supernovae  (SN Ia) \citep{riess98}, baryon acoustic oscillations (BAO) \citep{anderson12,beutler11}, as well as small scale anisotropies of the cosmic microwave background (CMB) radiation \citep{hinshaw13,planck13}. Being bright beacons from deep universe, gamma-ray bursts (GRBs) have been considered as a potential complementary probe to conduct cosmography. Many authors have made use of GRB luminosity indicators as standard candles \citep[e.g.][]{dai04,ghirlanda04,liangzhang05,schaefer07}. However, unlike the SN Ia candle, GRB correlations lack physical motivation and usually have relatively large scatter, so that their role as standard candles is debated. Nonetheless, GRBs can serve as a complementary tool to probe the relatively high-$z$ universe, such as the star formation history, the metal enrichment history, and the properties of intergalactic medium \citep[e.g.][]{barkana04,mcQuinn08,virgili11,wang12}.

Recently, \cite{thornton13} reported the discovery of a new type of cosmological transients, dubbed Fast Radio Bursts (FRBs). These objects have anomalously high dispersion measure (DM) values corresponding to a cosmological redshift between 0.5 and 1 \citep{lorimer07, thornton13}\footnote{Some recent works also invoke a galactic origin for some of the FRBs \citep{loeb14,bannister14,kulkarni14}.}. If the redshifts of these events can be measured, the combination of $z$ and DM information would be invaluable to conduct cosmography. \cite{zhang14} suggested that a small fraction of FRBs could be physically connected to some GRBs, whose central engine is a supra-massive millisecond magnetar which collapses to a black hole at $10^2 - 10^4$ seconds after the burst\footnote{Most FRBs would be produced by supra-massive NSs collapsing into a black hole after a much longer delay, e.g. thousands to millions of years \citep{falcke14}.}. Two possible such associations might have been observed by \cite{bannister12}, and the fraction of GRBs that might host a magnetar central engine, and hence, a possible FRB, could be up to $\sim 60\%$ for long GRBs \citep{lu14} and probably an even higher fraction for short GRBs \citep{rowlinson13}. Using the dispersion measure (DM) values of the two possible FRB/GRB association candidates \citep{bannister12}, \cite{deng14} derived the upper limits of the baryon mass density along the line-of-sight of the two GRBs, which are consistent with the values derived by other methods. This lends further support to FRB/GRB associations. The FRB/GRB association systems, if commonly detected in the future, would be an ideal tool to constrain cosmological parameters and properties of dark energy at redshifts not attainable by SN Ia. We term this prospect as ``FRB/GRB cosmography''.

In this paper, we study the prospects of conducting FRB/GRB cosmography in detail. An independent work was recently carried out by \cite{zhou14}, who discussed using FRBs to constrain dark energy properties assuming that the redshifts of FRBs can be measured.

\section{Uncertainties in intergalactic medium DM value estimation}

For an FRB/GRB system, the measured dispersion measure \citep{deng14}
\begin{equation}
{\rm DM}_{\rm  obs}={\rm DM}_{\rm MW}+{\rm DM}_{\rm IGM}+{\rm DM}_{\rm Host}+{\rm DM}_{\rm GRB}
\label{eq:DM}
\end{equation}
has contributions from the Milky Way, intergalactic medium, GRB host galaxy, and the GRB blastwave, respectively. Among these terms, 
\begin{eqnarray}
{\rm DM}_{\rm IGM} & = & {\rm DM}_{\rm  obs}-\left({\rm DM}_{\rm MW}+{\rm DM}_{\rm GRB}+{\rm DM}_{\rm Host}\right) \nonumber \\
& = & \left< {\rm DM}_{\rm IGM} \right> + \Delta ({\rm DM}_{\rm IGM})
\label{eq:DMs}
\end{eqnarray} 
is the relevant one to probe the universe.
Here $\left< {\rm DM}_{\rm IGM} \right> (z)$ is the IGM DM averaged in all directions for a given $z$, which is defined by cosmological parameters. By introducing the fraction of ionized electrons in hydrogen (H) and helium (He) atoms as a function of redshift ($\chi_{\rm e,H}(z)$ and $\chi_{\rm e,He}(z)$), and assuming $\rm {H:He}$ mass ratio is approximately $3:1$, one can give a general expression for $\left<{\rm DM}_{\rm IGM}\right>$ by generalizing Eq.(13) of \cite{deng14}:
\begin{eqnarray}
\left<{\rm DM}_{\rm IGM}(z)\right> & &= \frac{3cH_0\Omega_bf_{\rm IGM}}{8\pi G m_p} 
\int_{0}^{z} \frac{\chi(z')
(1+z')dz'}{E(z')}.
\label{eq:IGM}
\end{eqnarray}
where
\begin{eqnarray}
&&\chi(z)=\frac{3}{4}y_1\chi_{\rm e,H}(z)+\frac{1}{8}y_2\chi_{\rm e,He}(z), \nonumber \\
&&E(z)=[(1+z)^{3}\Omega_{\rm M}+f(z)\Omega_{\rm DE}+(1+z)^{2}\Omega_{\rm k}]^{1/2}, \nonumber\\
&&f(z)=\exp\left[3\int_{0}^{z}\frac{(1+w(z''))dz''}{(1+z'')}\right], \nonumber
\end{eqnarray}
$\Omega_b$ is the current baryon mass fraction of the universe, $f_{\rm IGM}$ is the fraction of baryon mass in the intergalactic medium, and $y_1 \sim 1$ and $y_2 \sim 1$ are IGM hydrogen and helium mass fractions normalized to 3/4 and 1/4, respectively. The term $\Delta ({\rm DM}_{\rm IGM})$ (can be both positive and negative) in Eq.(\ref{eq:DMs}) stands for deviation ${\rm DM}_{\rm IGM}$ from the mean value at individual lines of sight due to the inhomogeneity of the baryon matter in the universe \citep[e.g.][]{mcQuinn14}.

Many dark energy models invoke $w \neq -1$, and many have $w(z)$ not a constant. Ideally (if $\Delta {\rm DM}_{\rm IGM}$ is not large), these models may be differentiated with a good sample of FRB/GRB systems spreading in a wide redshift range, as long as one could precisely measure ${\rm DM}_{\rm obs}$ and precisely determine $\rm {DM_{\rm MW}+DM_{\rm Host}+DM_{\rm GRB}}$. 
The advantage of using $\left<{\rm DM}_{\rm IGM}\right>$ to conduct cosmography is that the underlying physics is clean, which stems from the simple geometry of the universe, in constrast to other standard candles or rulers that invoke messier physics (e.g. SN Ia candle relies on poorly known supernvoa explosion physics). In the following, we discuss how to practically determine various DM components and estimate their relevant uncertainties in turn.

\begin{itemize}
\item  The measurement of $\rm{DM_{obs}}$ is very accurate, for instance, the uncertainties for the four reported FRBs are 0.05, 0.3, 0.7 and 0.3 ${\rm pc~cm^{-3}}$ respectively. Here we use an average of these four values to estimate the uncertainty of $\rm{DM_{obs}}$, i.e, $\sigma_{\rm obs}=0.34~ {\rm pc~cm^{-3}}$, which is negligible compared with other uncertainties.
\item  $\rm {DM_{MW}}$ can be estimated to within a factor of 1.5-2 using Galactic pulsar data \citep{taylor93}.  It rapidly drops to small values 
as the Galactic latitude is $|b|>10^{\rm o}$. We therefore suggest to take this condition as our sample selection criterion in the future. 
%In this way, the uncertainty of $\rm{DM_{\rm MW}}$ should be small. 
With the ATNF pulsar data\footnote{http://www.atnf.csiro.au/research/pulsar/psrcat/} \citep{manchester05}, we find that the average dispersion of $\rm{DM_{\rm MW}}$ for $|b|>10^{\rm o}$ sources is 33 $\rm{pc~cm^{-3}}$ (see Fig. 1a), and we take this value as $\sigma_{\rm MW}$.
% and it should not be redshifted.
\item  In principle, one could precisely calculate $\rm{DM_{\rm GRB}}$ based on the GRB afterglow models if the model parameters could be constrained. \cite{deng14} presented some calculation results given typical parameters and found that $\rm{DM_{\rm GRB}}$ is typically around 1 $\rm{pc~cm^{-3}}$ for the ISM afterglow model and 10 $\rm{pc~cm^{-3}}$ for the wind afterglow model\footnote{Note that for extreme parameters in the wind model, $\rm{DM_{\rm GRB}}$ could reach 100 \citep{deng14}. However, such cases could be identified through afterglow modeling and dropped out from the sample.}. Based on these results, we cautiously adopt $\sigma_{\rm GRB} =$ 10 $\rm{pc~cm^{-3}}$.
\item The value of $\rm{DM_{\rm Host}}$ depends on many factors, such as the type of GRB host galaxy, the site of GRB in the host galaxy, the inclination angle of the disk with respect to line of sight, and so on. 
Based on the DM dispersion of Milky Way, one may expect the uncertainty of $\rm{DM_{\rm Host}}$ could be from tens to 
hundreds of $\rm{pc~cm^{-3}}$. Here we take $\sigma_{\rm Host}$ as a free parameter.
Note that both ${\sigma_{\rm GRB}}$ and ${\sigma_{\rm Host}}$ should be redshifted.
\item  The uncertainty $\Delta ({\rm DM}_{\rm IGM})$ due to inhomogeneity of the baryon matter in the IGM is an unknown parameter. 
Numerical simulations \citep{mcQuinn14} gave a standard deviation ${\sigma_{\rm IGM}} \sim 100-400 {\rm ~pc~ cm^{-3}}$ around the mean value $\left< {\rm DM_{\rm IGM}} (z)\right>$ at $ z = 0.5- 1$. If so, the IGM inhomogeneity effect would be the dominant component for $\rm{DM_{IGM}}$ uncertainty. 
Without any observational guide, we introduce an unspecified $\sigma_{\rm IGM}(z)$ to describe this uncertainty.
\end{itemize}
Given a certain set of cosmological model parameters ($H_0$, $\Omega_b$, $f_{\rm IGM}$, $\Omega_{\rm M}$, $\Omega_{\rm DE}$, $\Omega_{\rm k}$, $\chi_{\rm e,H} (z)$, and $\chi_{\rm He,H} (z)$), one can calculate $\left< {\rm DM}_{\rm IGM} \right>$ precisely. However, if one infers this value from the observed ${\rm DM}_{\rm obs}$ (Eq.(\ref{eq:DMs})), one would have to incorporate a total uncertainty of (in unit of ${\rm pc~cm^{-3}}$)
\begin{eqnarray}
\sigma_{\left<\rm DM_{IGM}\right>}&=&\sigma_{\rm obs}+\frac{\sigma_{\rm GRB}+\sigma_{\rm Host}}{1+z}+\sigma_{\rm MW}+\sigma_{\rm IGM}(z)\nonumber\\
&=&\frac{10+\sigma_{\rm Host}}{1+z}+33.34+\sigma_{\rm IGM}(z),
\label{eq:sigma}
\end{eqnarray}

\begin{figure}
\centering
\plotone{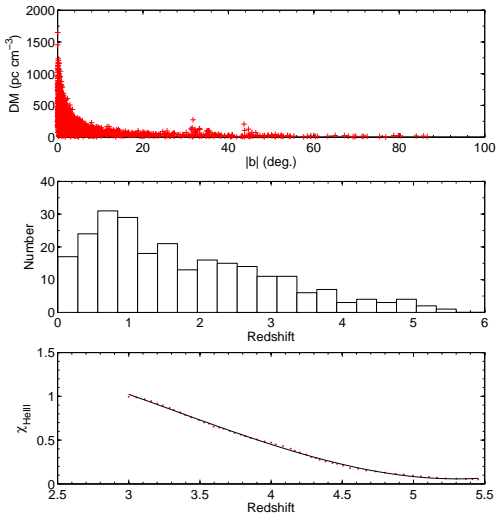}
\caption{a): Measured DM for known pulsars in Milky Way against their Galactic latitude; b) The observed GRB redshift distribution; c) The $\chi_{\rm HeIII}(z)$ evolution history from numerical simulations of \cite{mcQuinn09} (red dots, their D1 model) and our analytical approximation (solid line). }
\label{fig:compre}
\end{figure}

\section{Testing capability of FRB/GRB systems to conduct cosmography}

We perform Monte Carlo simulations to test how well FRB/GRB systems can be used to constrain the dark energy equation of state. To do so, we need to assume an underlying cosmological model (i.e. effectively fix a set of cosmological parameters), and then simulate a sample of FRB/GRB systems each with an assigned $z$ and $\left<{\rm DM}_{\rm IGM} \right>(z)$. For the $z$-distribution, since no observed FRB/GRB system could be used as a reference\footnote{The two reported candidate FRB/GRB systems \citep{bannister12} unfortunately did not have redshift measurements.}, we simulate the $z$-distribution of our sample based on the observed $z$ distribution of the observed GRBs\footnote{The data was collected from an online catalog listed at http://lyra.berkeley.edu/grbox/grbox.php.} (see Fig.\ref{fig:compre}b). For each GRB with an assigned $z$, we calculate its $\left<{\rm DM}_{\rm IGM} \right>(z)$ based on Eq.(\ref{eq:IGM}), and then assign a ${\rm DM}_{\rm IGM}$ value through introducing the scatter $\sigma_{\left< {\rm DM}_{\rm IGM} \right>}$ defined by Eq.(\ref{eq:sigma}). 

To make use of Eq.(\ref{eq:IGM}), we need to assign certain values to relevant parameters. Since $\rm{DM_{\rm IGM}}$ is linearly proportional to $H_0$, $\Omega_b$ and $f_{\rm IGM}$, these three parameters have to be constrained independently for our purpose. Incidentally, these parameters can be constrained independent of the dark energy models. The Hubble constant $H_0$ can be constrained using the conventional extragalactic distance scale, while $\Omega_b$ can be constrained by CMB or Big Bang nucleosynthesis (BBN) data.  Here we adopt the following ``benchmark'' values recently derived from the joint $Plank+WMAP$ data \citep{hinshaw13,planck13} in our simulations: $H_0=(67.3 \pm 1.2)~ {\rm km~s^{-1}~Mpc^{-1}}$ and $\Omega_b=0.0487\pm0.002$.
The value of $f_{\rm IGM}$ is more uncertain. According to the baryon mass summation results of \cite{fukugita98}, one could derive an estimation of $f_{\rm IGM}\sim 0.83$ \citep{deng14}. Recent simulation results show that for redshifts $z\leq 0.4$,  the collapsed phase (galaxies, groups, clusters, etc.) contains $18\% \pm 4\%$ baryon mass, which gives $f_{\rm IGM}\sim 0.82\pm0.04$ \citep{shull12}. Here we suggest to adopt  $f_{\rm IGM}\sim 0.83$ as the prior.
In principle, the mean value of the product $\Omega_b f_{\rm IGM}$ could also be directly measured with a large sample of nearby FRB/GRB systems in the future \citep{deng14}.

 It should be safe to assume $\chi_{\rm e,H}(z)=\chi_{\rm e,He}(z)=1$ for nearby FRB/GRB systems at $z<3$,  since both H and He are expected to be fully ionized \citep{fan06,mcQuinn09}. However, to test dark energy equation of state and its dynamical evolution of dark energy $w(z)$, samples with larger redshifts are essential. More accurate expressions for $\chi_{\rm e,H}(z)$ and $\chi_{\rm e,He}(z)$ are required to reduce the uncertainty. At $z > 6$ hydrogen reionization becomes important while the reionization history is poorly known. {\em We suggest that in the future one should use FRB/GRB systems at $z < 6$ to perform cosmography studies}. In this redshift range, one can approximately take $\chi_{\rm e,H}(z)=1$. 
The fraction of electrons in He atoms that has been ionized should be $\chi_{\rm e,He}(z) = (1/2)\chi_{\rm HeII}(z)+\chi_{\rm HeIII}(z)=(1/2)(1+\chi_{\rm HeIII}(z))$, since the ionization energy of HeI is close to that of H, one may assume that HeI is also fully ionized. \cite{mcQuinn09} have studied HeII ionization history through detailed simulations. For easy application, we fit their numerical results (D1 model in \cite{mcQuinn09}) with a polynomial, so that the mean value of $\chi_{\rm e,He}(z)$ can be approximated as
\begin{eqnarray}
\chi_{\rm e,He}(z)=\left\{ \begin{array}{ll} 1, & z<3;\\
0.025z^3-0.244z^2+0.513z+1.006, &
z>3; \\
\end{array} \right.
\label{eq:chihe}
\end{eqnarray}
Such an analytical approximation has a $\sim 4\%$ error with respect to the numerical results (Fig.1c), and the simulation results are also slightly model-dependent. Nonetheless, in view of the 1/8 coefficient of $\chi_{\rm e,He}(z)$ in Eq.(\ref{eq:IGM}), these uncertainties are negligible for our purpose.

After fixing the above parameters, we want to test how simulated mock data constrain the underlying dark energy models. We assume an underlying flat $\Lambda$CDM model with $\Omega_{\rm M}=0.315$ and $\Omega_\Lambda = 0.685$, but introduce a general set of $w$CDM models and apply the mock data to check how well the data can reproduce the underlying model.

\begin{figure}
\centering
\plotone{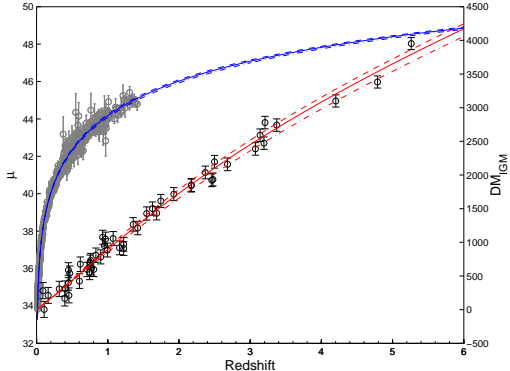}
\caption{Hubble Diagram and $\left<{\rm DM}_{\rm IGM}\right>$ evolution for different $w$CDM models (a flat universe is assumed here and $\Omega_m$ is fixed at 0.315). The solid lines are for $w=-1$, dash lines are for $w=-1.2$, and the dash-dot lines are for $w=-0.8$.  Grey circles are for Union 2.1 SN Ia sample \citep{suzuki12}, and black circles show one example of our simulated FRB/GRB sample ($\rm {N_{sam}}=50$, $\sigma_{\rm Host}=30 \rm{pc~cm^{-3}}$ and $\sigma_{\rm IGM}=50 \rm{pc~cm^{-3}}$). }
\label{fig:DMDL}
\end{figure}

Figure \ref{fig:DMDL} shows the theoretical $\left<{\rm DM}_{\rm IGM}\right>$ (red curves) and distance modulus $\mu$ (blue curves) as a function of $z$ for three $w$CDM models ($\Omega_{\rm k}=0$, $w$ is constant, not evolving with $z$): $w=-1$ (solid), $w=-1.2$ (dashed), and $w=-0.8$ (dash-dotted). Overplotted are the simulated 50 FRB/GRB systems and the observed Union 2.1 SN Ia sample \citep{suzuki12}. From this plot, it is clearly seen that the DM curves have a wider separation than the $\mu$ curves to allow an easier differentiation among the models. This is especially so at high redshifts. While all SN Ia are at $z<2$, GRBs have been detected at redshifts as high as $z=8.2$ \citep{tanvir09,salvaterra09}. Since FRBs are typically bright, with a peak flux at multi-Jansky level. A moderately large radio telescope with rapid slewing capability would lead to detection of FRBs following GRBs in the redshift range $2 < z < 6$. As a result, FRB/GRB systems may be a viable way to constrain dark energy equation of state.

\begin{figure}
\begin{minipage}[b]{0.5\textwidth}
\centering
\includegraphics[width=2.7in]{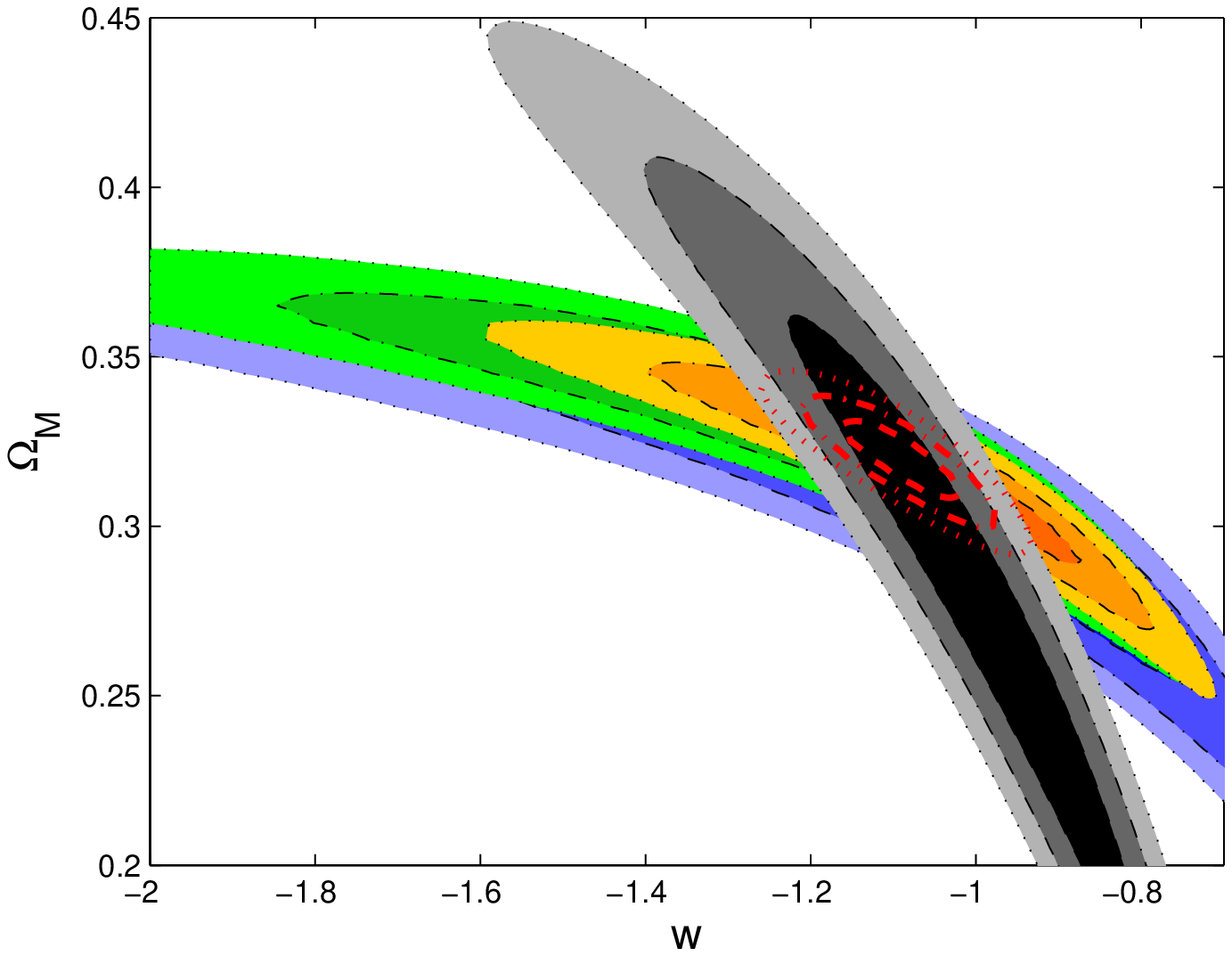}
\end{minipage} \\%
\begin{minipage}[b]{0.5\textwidth}
\centering
\includegraphics[width=2.7in]{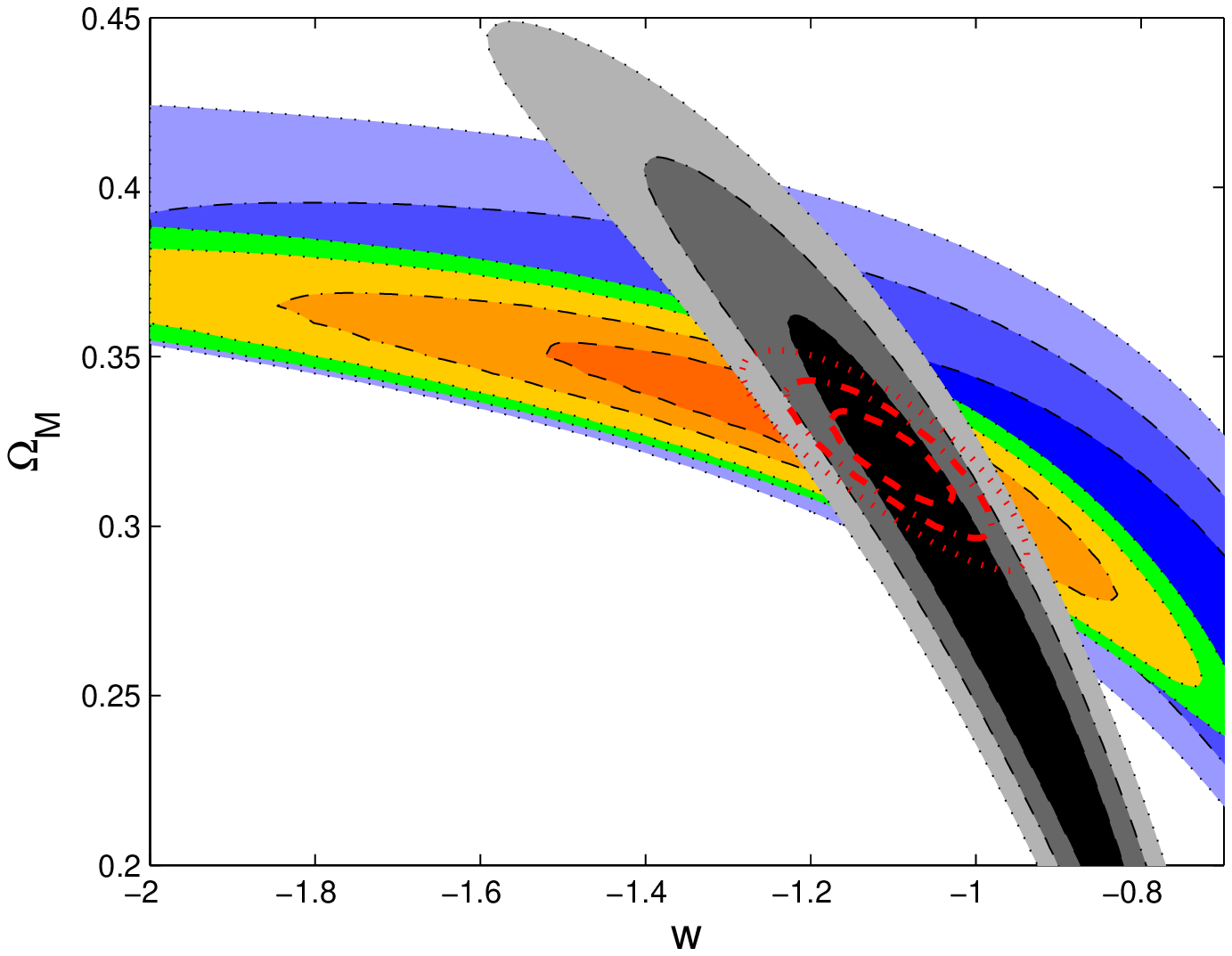}
\end{minipage} \\%
\caption{Constraint results for $w$CDM models using simulated FRB/GRB systems, compared with Union 2.1 SN Ia results (grey contours): (a) effect of different sample sizes: ${N_{\rm sam}=30}$ (blue contours), 60 (green contours), and 100 (orange contours). Other parameters are fixed to $\sigma_{\rm Host}=30,\sigma_{\rm IGM}=50$;
(b) effect of $\sigma_{\rm Host}$ and $\sigma_{\rm IGM}$. $N_{\rm sam}$ is fixed to 60:
$\sigma_{\rm Host}=30,\sigma_{\rm IGM}=200$ (blue contours), $\sigma_{\rm Host}=100,\sigma_{\rm IGM}=50$ (green contours), and $\sigma_{\rm Host}=30,\sigma_{\rm IGM}=50$ (orange contours).}
\label{fig:sample}
\end{figure}

To see this point more clearly, we show contour constraints in the $\Omega_{\rm M} - w$ 2-dimensional plane. For simplicity, we take $\sigma_{\rm IGM}(z)$ as a constant. We first fix the uncertainties of host galaxy and IGM inhomogeneity as $(\sigma_{\rm Host},\sigma_{\rm IGM})=(30,50)$, and generate three samples with $\rm{N_{sam}}=30,~60,~100$, respectively\footnote{These sample sizes are adopted according to a realistic estimate of possible FRB/GRB systems that might be detected in a decade time scale, based on the detection rate derived from true searches \citep{bannister12} and GRB data analysis \citep{lu14}.}. Figure \ref{fig:sample}a shows the contour contrast for different sample sizes. For comparison, we also plot the contours by applying the Union 2.1 SN Ia data. One can see that with a moderate sample size of several 10s, the contour size of FRB/GRB systems is already comparable to that of SN Ia (which has more than 500 SN Ia). Furthermore, when combining FRB/GRB systems with SN Ia, a much better constraint is achieved. This is mainly due to the fact that FRB/GRB systems are distributed in a much wider redshift range towards high-$z$, where better constraints on the models can be achieved.
Our results are generally consistent with \cite{zhou14}.

We also test the effects from uncertainties of the host galaxy and inhomogeneity. In this case, we fix our sample number as 60 and generate three samples with $(\sigma_{\rm Host},\sigma_{\rm IGM})=(30,50),~(30,200),~(100,50)$, respectively. As shown in Figure \ref{fig:sample}b, the results are more sensitive to $\sigma_{\rm IGM}$ than $\sigma_{\rm Host}$. This is because the latter becomes less significant at high redshifts due to the $(1+z)$ factor, while the high redshift data are more powerful to differentiate among the models.

\section{$D_{\rm L}(z)$ vs. $\left<{\rm DM}_{\rm IGM}\right>(z)$ and implication}

\begin{figure}
\centering
\plotone{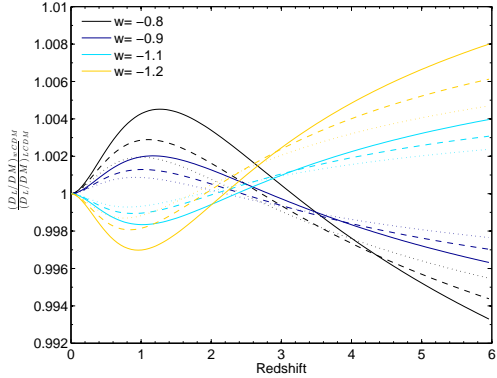}
\caption{The ${D_{\rm L}(z)/\left<{\rm DM_{IGM}}\right>(z)}$ ratio in different $w$CDM models ($w=-0.8, -0.9, -1.1, -1.2$) normalized to that of the $\Lambda$CDM model. For each model, three $\Omega_{\rm M}$ values are adopted: 0.2 (solid), 0.3 (dashed), 0.4 (dotted).}
\label{fig:ratio}
\end{figure}

``Standard candles'' make use of the luminosity distance of the source, $D_{\rm L}(z)=cH_0^{-1}(1+z) \int_{0}^{\rm z} \frac{dz'}{E(z')}$.  Comparing with Eq.(\ref{eq:IGM}), one has
\begin{eqnarray}
\frac{D_{\rm L}(z)}{\left<{{\rm DM}_{\rm IGM}}\right>(z)}=\frac{8\pi G m_p}{3H_0^2\Omega_bf_{\rm IGM}} \frac{(1+z) \int_{0}^{z} \frac{dz'}{E(z')}}{\int_{0}^{z} \frac{
\chi(z')(1+z')dz'}{E(z')}}.
\end{eqnarray}
For given $H_{0},\Omega_b$ and $f_{\rm IGM}$ values (measured independently), different dark energy cosmology models are contained in the expression of $E(z)$, which is essentially canceled out in the $D_{\rm L}(z)/\left<{\rm DM}_{\rm IGM} \right> (z)$ ratio. This ratio is expected to only weakly depend on dark energy models. To verify this, we again take the $w$CDM cosmology models as an example. We choose a relatively large parameter space with $-1.2<w<-0.8$ and $0.2<\Omega_{\rm M}<0.4$. As shown in Figure \ref{fig:ratio}, the ${D_{\rm L}/{\rm DM_{IGM}}}$ ratio only differs by less than 1\% with respect to the $\Lambda$CDM model at $z<6$, which is indeed negligible. 

This insensitivity of the $D_{\rm L} (z)/ \left<{\rm DM}_{\rm IGM}\right> (z)$ ratio on the dark energy models makes it convenient to combine standard candles (e.g. Type Ia SNe) and FRB/GRB pairs to conduct cosmography. While the theoretical values of $\left<{\rm DM}_{\rm IGM}\right> (z)$ are well defined, determining them from the data is not easy, which requires to accumulate a large enough sample of FRB/GRB samples in many redshift bins to cancel out the inhomogeneity effect from different lines of sight. The insensitivity of the $D_{\rm L} (z)/ \left<{\rm DM}_{\rm IGM}\right> (z)$ ratio allows one to easily determine the {\em shape} of $\left<{\rm DM}_{\rm IGM}\right> (z)$ based on the well-mapped $D_{\rm L} (z)$ from the SN Ia data (regardless of the dark energy models). Even though the normalization of $\left<{\rm DM}_{\rm IGM}\right> (z)$ depends on $H_0$, $\Omega_b$, and $f_{\rm IGM}$, by knowing the shape of $\left<{\rm DM}_{\rm IGM}\right> (z)$ one can combine FRB/GRB systems at all redshifts to ``calibrate'' $\left<{\rm DM}_{\rm IGM}\right> (z)$ and find out the normalization. This requires a much smaller sample to achieve the calibration purpose. With  $\left<{\rm DM}_{\rm IGM}\right> (z)$ well mapped, one can then directly study the scatter of ${\rm DM}_{\rm ICM}$ due to local IGM inhomogeneity \citep[e.g.][]{mcQuinn14} as well as its redshift evolution (i.e. $\sigma_{\rm IGM} (z)$).

\section{Summary and discussion}

FRB/GRB systems, if confirmed to be commonly exist, have great potential to infer cosmological parameters, especially to constrain the equation of state of dark energy. We have shown that with a moderate sample size of several 10s, one may reach a constraint on $w$ comparable to a large SN Ia sample. Combining SN Ia data and FRB/GRB systems, one may achieve unprecedented accuracy in constraining $w$. The insensitivity of the $D_{\rm L}(z) / \left<{\rm DM}_{\rm IGM}\right> (z)$ ratio offers the advantage of using a relatively small sample to cabibrate $\left<{\rm DM}_{\rm IGM}\right> (z)$ and to diagnose the local IGM inhomogeneity as well as its redshift distribution.

Our method is also applicable if the redshifts of FRBs can be measured independently with other methods. \cite{zhou14} explored this possibility and reached the similar conclusion that FRBs can constrain dark energy equation of state and IGM inhomogeneity. FRB/GRB associations provide a practical method to measure redshifts of FRBs, and the sample size adopted in this paper is based on observational and theoretical insights of GRBs as well as the results of preliminary searches of FRB/GRB associations \citep{bannister12}.

In the end, we want to point out some caveats of our method:

1) If most of the FRBs (or even all of them) are eventually proved to be of a galactic origin, or the redshifts of FRBs could never be measured, FRBs are no longer relevant for cosmography.

2) As shown in our simulations, the constraint results for cosmological parameters are most sensitive to the intrinsic IGM inhomogeneity $\sigma_{\rm IGM}$. Even though the calibration of $\left<{\rm DM}_{\rm IGM}\right>$ could be achieved (see section 4), if $\sigma_{\rm IGM}$ is very large, its uncertainty could still reduce the constraint accuracy dramatically. A (much) larger sample size than simulated here is needed to make this method competitive.

3) The GRB hosts likely evolve with redshift \citep[e.g.][]{perley13}. This suggests that both $\rm{DM_{host}}$ and $\rm{\sigma_{host}}$ could be redshift dependent. In our simulations, we adopt a constant free parameter for $\rm{\sigma_{host}}$, and its contribution to $\sigma_{\rm DM_{IGM}}$ is reduced at higher redshifts due to $(1+z)$ factor (Eq.(\ref{eq:sigma}). If it turns out that $\rm{DM_{host}}$ and $\rm{\sigma_{host}}$ would increase with redshift and $\rm{\sigma_{host}}$ could become comparable or even larger than $\sigma_{\rm IGM}(z)$, the accuracy of our results could be reduced dramatically. To keep the method competitive, either a larger sample is required, or more observational information about the GRB host galaxy population is needed.

4) If the dispersions of various parameters is intrinsically non-Gaussian, a larger sample than simulated may be needed to quantify these dispersions to achieve an accurate constraint on dark energy models.

%\textbf{4) The redshift distribution for FRB/GRB systems could be biased by the redshift distribution of GRBs, which has a preferred redshift range from 1 to 3 (as shown in Fig. \ref{fig:compre}b ). On one hand, such a bias is helpful for determining $\rm{DM_{host}}$, $\rm{\sigma_{host}}$, $\left<{\rm DM}_{\rm IGM}\right>$ and $\rm{\sigma_{IGM}}$ within the particular redshfit range. On the other hand, a high concentration of redshift distribution may reduce the power of FRB/GRB systems to conduct cosmography.}

%\acknowledgment
\vskip 0.2in

We thank Fa-Yin Wang for a helpful discussion. This work is partially supported by National Basic Research Program (``973" Program) of China under Grant No. 2014CB845800, the NSFC (11273005) and SRFDP (20120001110064).  HG acknowledges a Fellowship from China Scholarship Program for support.

\end{document}